\documentstyle[twocolumn,aps,prl,epsf,amssymb]{revtex}

\begin{document}
\draft
\bibliographystyle{srt}
\def\vec#1{{\bf{#1}}}

\wideabs{
\title{
		Investigation of the Exclusive
		$^{3}{\text{He}}(e,e^{\prime}pp)n$ Reaction}

\author{
	D.L.\ Groep$^{1}$, 
	M.F. van\ Batenburg$^{1}$, 
	Th.S.\ Bauer$^{1,2}$, 
	H.P.\ Blok$^{1,3}$, 
	D.J.\ Boersma$^{1,2}$, 
	E.\ Cisbani$^{4}$, 
	R.\ De Leo$^{5}$, 
	S.\ Frullani$^{4}$, 
	F.\ Garibaldi$^{4}$, 
	W.\ Gl\"ockle$^{6}$, 
	J.\ Golak$^{7}$, 
	P.\ Heimberg$^{1,3}$, 
	W.H.A.\ Hesselink$^{1,3}$, 
	M.\ Iodice$^{4}$, 
	D.G.\ Ireland$^{8}$, 
	E.\ Jans$^{1,}$\cite{co}, 
	H.\ Kamada$^{6}$, 
	L.\ Lapik\'as$^{1}$, 
	G.J.\ Lolos$^{9}$, 
	R.\ Perrino$^{10}$, 
	A.\ Scott$^{8}$, 
	R.\ Starink$^{1,3}$, 
	M.F.M.\ Steenbakkers$^{1,3}$, 
	G.M.\ Urciuoli$^{4}$, 
	H. de\ Vries$^{1}$, 
	L.B.\ Weinstein$^{11}$, 
	H.\ Wita{\l}a$^{7}$
}

\address{$^{1}$NIKHEF, P.O.\ Box\ 41882, 1009\ DB\ Amsterdam, The\ Netherlands}
\address{$^{2}$Universiteit\ Utrecht, P.O.\ Box\ 80.000, 3508\ TA\ Utrecht,
	The\ Netherlands}
\address{$^{3}$Vrije\ Universiteit, de\ Boelelaan 1081, 1081\ HV\ Amsterdam,
	The\ Netherlands}
\address{$^{4}$Istituto\ Superiore\ di\
	Sanita\makebox[0pt]{\hspace*{3pt}\`\ }, Laboratorio\ di\ Fisica, INFN,
	Viale\ Regina\ Elena\ 299, Rome, Italy}
\address{$^{5}$INFN\ Sezione\ di\ Bari, Dipartimento\ Interateneo\
	di\ Fisica, Via\ Amendola\ 173, Bari, Italy}
\address{$^{6}$Institut\ f\"ur\ Theoretische\ Physik\ II,
	Ruhr-Universit\"at\ Bochum, D-44780\ Bochum, Germany}
\address{$^{7}$Institute of Physics, Jagellonian University, 
	PL-30059 Cracow, Poland}
\address{$^{8}$Department\ of\ Physics\ and\ Astronomy, University\
	of\ Glasgow, Glasgow G12 8QQ, UK}
\address{$^{9}$Department\ of\ Physics, University\ of\ Regina, Regina
	SK\ S4S\ 0A2, Canada}
\address{$^{10}$INFN\ Sezione\ di\ Lecce, via per Arnesano, 73100 Lecce, Italy}
\address{$^{11}$Physics\ Department, Old\ Dominion\ University, Norfolk,
	Virginia\ 23529}

\date{\today}

\maketitle
\begin{abstract}
	Cross sections for the $^{\text{3}}{\text{He}}
	(e,e^{\prime}pp)n$ reaction were measured at an energy transfer of
	220\ MeV and three-momentum transfers $q$ of 305, 375, and
	445\ MeV/$c$.  Results are presented as a function of $q$ and the
	final-state neutron momentum for slices in specific kinematic
	variables.  At low neutron momenta, comparison of the data
	to results of continuum Faddeev calculations performed with the
	Bonn-B nucleon-nucleon potential indicates a dominant role for
	two-proton knockout induced by a one-body hadronic current.
\end{abstract}

%
}

The ground-state properties of few-nucleon systems are an excellent
testing ground for models of the nucleon-nucleon (\emph{NN})
interaction.  Calculations of the structure of few-nucleon
systems have successfully been performed with realistic \emph{NN}
potentials, both phenomenological ones and those based on the exchange
of mesons and including a phenomenological description of the
short-range \emph{NN} behaviour\ \cite{nog97,car98}.  The parameters of
these models were fitted to the phase shifts obtained from \emph{NN}
scattering data.  Significant advances in solving the three-nucleon
(\emph{3N}) continuum\ \cite{glo96} currently allow exact calculations
of electron-induced nuclear reactions leading to the breakup of the
tri-nucleon system\ \cite{Gol95,Mei92}.  The exclusive
$^3\text{He}(e,e^{\prime}NN)N$ reaction is sensitive to details of the
initial nuclear state, as well as to the reaction mechanism and
rescattering effects in the final state.  The three-fold coincidence
experiments necessary to measure the cross sections for these
reactions have recently proven feasible\ \cite{Ond97,Ros95}. 

At intermediate electron energies, the cross section for
electron-induced two-nucleon knockout is driven by several processes. 
The \emph{NN} interaction induces initial-state correlations between
nucleons and therefore the coupling of a virtual photon to one of
these nucleons via a one-body hadronic current can lead to knockout of
both nucleons.  The interaction of the virtual photon with two-body
currents, either via coupling to mesons or via intermediate $\Delta$
excitation, will also contribute to the cross section.  In addition,
final-state interactions\ (FSI) among the nucleons after absorption of
the virtual photon may cause breakup of the tri-nucleon system.  By
studying the dependence of the cross section on various kinematic
quantities one may hope to unravel the tightly connected properties of
the \emph{NN} interaction, short-range correlations, two-body currents
and FSI.  In this Letter we present results of a
$^{3}\text{He}(e,e^{\prime}pp)n$ experiment that was performed in the
dip region, i.e., the kinematic domain in between the peaks due
to quasi-elastic scattering and $\Delta$ excitation.

The measurements were performed with the high duty-factor electron
beam extracted from the Amsterdam Pulse Stretcher ring at NIKHEF.  The
incident electrons had an energy of 564\ MeV.  A cryogenic,
high-pressure barrel cell containing gaseous $^{\text{3}}\text{He}$
was used.  The luminosity amounted to $5\times10^{35}$ atoms
cm$^{-2}$s$^{-1}$.  The scattered electrons were detected in the QDQ
magnetic spectrometer\ \cite{Vri84} and the emitted protons in highly
segmented plastic scintillator arrays\ \cite{Pel99}.  The kinematic
settings of the QDQ correspond to a virtual-photon energy of
$\omega$=220\ MeV and three-momentum transfer values of $q$=305, 375,
and 445\ MeV/$c$.  At $q$=305\ MeV/$c$, protons were detected in the
angular ranges $5^{\circ}<\gamma_1<60^{\circ}$ and
$-170^{\circ}<\gamma_2<-110^{\circ}$, where $\gamma_i$ is the angle
between the momentum $\vec{p}^\prime_i$ of proton $i$ and the
transferred three-momentum $\vec{q}$.  For $q$=375 and 445\ MeV/$c$,
$\gamma_1$ ranges between 5$^{\circ}$ and 30$^{\circ}$.  The
acceptance of the second scintillator array is $\pm 20^{\circ}$ both
in-plane and out-of-plane.  The central value of $\gamma_2$ was chosen
such that each configuration includes the point for which the neutron
is left at rest.  The detection threshold of the proton detectors was
72 and 48\ MeV for the kinetic energy of the protons emitted in forward
and backward direction, respectively.

Accidental coincidences were subtracted from the measured yield via a
procedure described in Ref.\ \cite{Ond98}.  The data were corrected for
electronics dead time and for inefficiencies due to multiple
scattering and inelastic processes of the protons in the detection
systems.  
Cross sections were obtained
by normalizing the yield to the detection volume and integrated
luminosity.
The results are presented as a function of the
missing momentum $|\vec{p}_m| =
|\vec{q}-\vec{p}^\prime_1-\vec{p}^\prime_2|$, of $\gamma_1$ and of
$p_{13}$, defined as $p_{13}= |\vec{p}^\prime_1-\vec{p}^\prime_3|$.

At missing energies $E_m= \omega- T_{p_1^\prime}- T_{p_2^\prime}-
T_{p_3^\prime}$ below the pion production threshold the kinematics of
the $^{\text{3}}\text{He}(e,e^{\prime}pp)n$ reaction is completely
determined, as $\vec{p}'_3\equiv \vec{p}_m$.  The measured
missing-energy spectrum contains a peak, corresponding to the
three-body breakup of $^{3}{\text{He}}$, which has a width of
6\ MeV\ (FWHM) due to resolutions of the detectors.  Strength has been
shifted from this peak towards higher missing energies due to
radiative processes.  The nine-fold differential cross section was
integrated over a range in excitation energy up to 14\ MeV, taking into
account the Jacobian $\partial T_{p_2^\prime} / \partial E_m$.  The
strength beyond this cutoff was estimated with a formalism similar to
that for the $(e,e^{\prime})$ reaction\ \cite{MoT69} and applied as an
overall correction factor to the data. 

In all figures only the statistical errors are indicated.  The
systematic error on the cross sections is 7\%.  It is mainly
determined by the uncertainty in the integrated luminosity\ (3\%), the
uncertainty in the correction applied for hadronic interactions and
multiple scattering in the proton detectors\ (6\%), and the
determination of the electronics dead time\ (2\%).

The data are compared to results of Faddeev calculations\ \cite{Gol95},
where both the three-nucleon bound-state and final-state wave
functions are exact solutions of the \emph{3N} Faddeev equations solved
in a partial-wave decomposition using the Bonn-B \emph{NN}
interaction.  By including the rescattering contributions to all
orders in the continuum, final state interaction effects are
taken into account completely.  To ensure convergence of the
calculations, \emph{NN} force components are included up to two-body
angular momenta $j=3$.

Two types of calculations were performed.  The first one only employs 
a one-body hadronic current operator.  The other one also includes 
processes, in which the virtual photon interacts with a $\pi$ or
$\rho$ meson (MECs), either in-flight or in a nucleon-meson vertex. 
Hence, the hadronic current operator is augmented with additional
currents $\vec{j}_{\pi}$ and $\vec{j}_{\rho}$ as proposed by
Schiavilla \emph{et al.}\ \cite{Sch89}.  At the vertices, cutoff
parameters $\Lambda_{\pi}=1.7\ \text{GeV}$ and
$\Lambda_{\rho}=1.85\ \text{GeV}$ have been introduced for the $\pi$
and $\rho$ meson-exchange interactions to take the finite size of
baryons and mesons into account.  In the case of direct two-proton
knockout, the contribution of MECs to the cross section is
suppressed, as the virtual photon --\ to first relativistic order\ --
cannot interact with a neutral meson exchanged in a \emph{pp}
system.

\begin{figure}
    \epsfxsize=3.375in
    \centerline{\epsffile{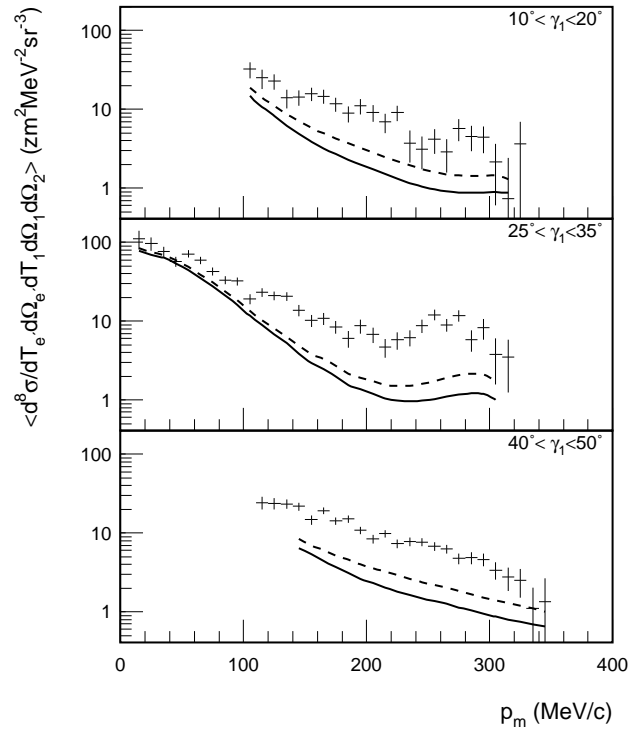}}
    \caption{
		The averaged eight-fold differential cross section for the reaction
		$^{3}\text{He}(e,e^\prime pp)$ for the kinematic setting
		$(\omega,q)$=$(220\ {\text{MeV}}, 305\ {\text{MeV}}/c)$ as a
		function of the missing momentum.  The data were averaged over
		the $\gamma_2$ range from 114$^{\circ}$ to 142$^{\circ}$ for
		three slices of $\gamma_1$.  The solid (dashed) curves
		represent the results of continuum Faddeev calculations
		without (including) MECs.
		\label{fig:pm}
	}
\end{figure}

The calculation of the differential cross section was performed in two
stages.  The process of solving the Faddeev equations, which is
computationally the most involved, was performed once per given
$(\omega,q)$ setting.  Cross sections could subsequently be calculated
for specific three-nucleon final states, determined by $\vec{p}'_1$ and
$\vec{p}'_2$.  This is especially important when comparing theoretical
predictions to data obtained with large-acceptance detectors, since
the cross section varies significantly within the region of phase
space covered by the experimental detection volume.  An orthogonal
grid in the laboratory quantities
$(\theta_1,\phi_1,\theta_2,\phi_2,T_1)$ was employed to compute the
cross section in each bin by averaging over the contributing part of
the experimental phase space.  A sufficiently large number of grid
points ($2.5\times10^6$ points per setting) was taken to ensure
convergence within 5\%.

The cross sections measured at $q$=305\ MeV/$c$ are shown in
Fig.\ \ref{fig:pm} as a function of the neutron momentum
($p^\prime_3\equiv p_m$) in the final state.  The three panels
correspond to different ranges in $\gamma_1$.  The largest range in
$p_m$ is spanned for $25^{\circ}<\gamma_1<35^{\circ}$.  Here, the
cross section was determined down to $p_m$ values as low as
10\ MeV/$c$.  In all panels the cross sections decrease roughly
exponentially with approximately the same slope as a function of
$p_m$.  This decrease reflects the neutron momentum distribution
inside $^{\text{3}}{\text{He}}$, for a specific range of $pp$ relative
momenta.  In the measured configuration the domain of relative momenta
around 300\ MeV/$c$ per nucleon was probed.  The curves show the
results of continuum Faddeev calculations.  The solid curve represents
results obtained with a one-body hadronic current only, while the
dashed line also includes contributions due to MECs.  As can be seen in
the middle panel, for $p_m \lesssim 100\ \text{MeV}/c$ there is a fair
description of the data by the solid curve and the contribution due to
MECs is almost negligible.

At higher missing momenta calculations including only one-body
currents fall short by about a factor of five.  The discrepancy between
data and calculations is likely to be due to two-body hadronic processes,
which involve coupling of the virtual photon to a proton-neutron
system in the initial state, thus generating neutrons with a large
momentum in the final state.  This is reflected in the
increase of the MEC contribution from 5\% to 40\% of the calculated cross
section towards high $p_m$, which is clearly not enough to explain the
discrepancy with the data.

Also excitation of the $\Delta$ resonance followed by its
non-mesonic decay has to be considered. The contribution of this
process strongly depends on the invariant mass $W_{\gamma NN}$ of
the virtual photon plus two-nucleon system.  In a direct reaction
mechanism, i.e., at low $p_m$, this is equal to the invariant
mass of the two-proton system in the final state, $W_{p^\prime_1
p^\prime_2}$.  For all kinematic settings we have $2030 \lesssim
W_{p^\prime_1 p^\prime_2} \lesssim 2055\ \text{MeV}/c^2$, which is well
below the mass of the $N\Delta$ system.  Moreover, in the two-proton
case the contribution of $pp\rightarrow \Delta^{+}p \rightarrow pp$
will be further suppressed because of angular-momentum and parity
conservation selection rules.  At high $p_m$ intermediate $\Delta$
excitation via the absorption of the virtual photon by a $pn$ pair
will contribute substantially to the reaction.  This process is known
to dominate the $(\gamma,pn)$ reaction at $E_\gamma >
180\ \text{MeV}$\ \cite{bof93}.  The invariant mass $W_{p^\prime
n^\prime}$ of a $pn$ system at $p_m \approx 300\ {\text{MeV}}/c$ is
around 2130\ MeV/$c^2$, which is close to the resonant mass of the
$N\Delta$ system.

It should be noted that --\ for a fixed value of $p_m$\ -- the angle
$\gamma_1$ implicitly defines the kinematic configuration of the final
state, provided that the direction of $\vec{p}^\prime_2$ is kept
within a limited range.  For $\gamma_1\lesssim 25^{\circ}$ and
$\gamma_1\gtrsim 35^{\circ}$, the three nucleons are always emitted at
sizeable angles with respect to each other, which reduces their mutual
interactions.  Enhanced probability for rescattering occurs in
so-called `FSI configurations' where two nucleons are emitted with
approximately the same momentum and angle.  Within the interval
$25^{\circ}<\gamma_1<35^{\circ}$, an `FSI configuration' between
proton-1 and the undetected neutron occurs around
$p_m=300\ \text{MeV}/c$, which introduces the bump observed in the data
for $200 \lesssim p_m \lesssim 300\ \text{MeV}/c$.  A similar structure
is seen in the calculated curves.

The enhanced rescattering effects occuring around `FSI configurations'
are the dominant factor that determines the cross section in these
regions.  The occurrence of such configurations can not be avoided in
experiments with large-solid-angle detectors.  On the other hand,
their presence offers the possibility to test the treatment of FSI
in calculations.  For these kinematic conditions, the cross section is
best represented as a function of the momentum difference between the
two outgoing nucleons involved:  $p_{ij}= |\vec{p}^\prime_i-
\vec{p}^\prime_j|$.  Here $p_{ij}= 0\ \text{MeV}/c$ corresponds to the
actual `FSI configuration'.

\begin{figure}
    \epsfxsize=3.375in
    \centerline{\epsffile{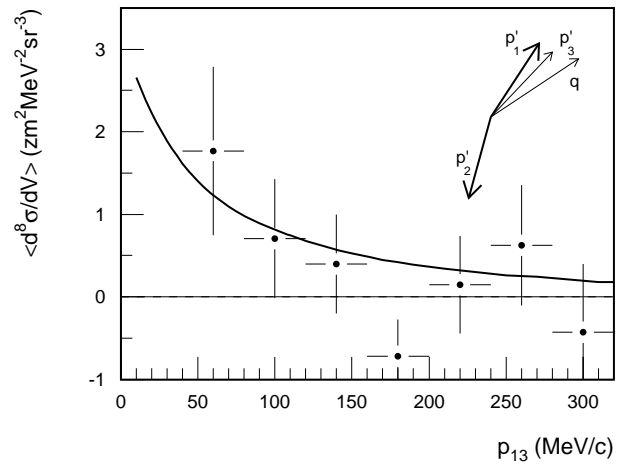}}
    \caption{
		Average cross section as a function of the momentum difference
		$p_{13}$ for the proton-neutron `FSI configuration' in the
		$(\omega,q)=(220\ {\text{MeV}},445\ {\text{MeV}}/c)$ kinematic
		setting.  The solid curve represents the continuum Faddeev
		result obtained with a one-body hadronic current. A
		schematic representation of the kinematics is indicated.
		\label{fig:p13}
	}
\end{figure}

As mentioned before, the influence of such an `FSI configuration' is
already apparent in Fig.\ \ref{fig:pm} at $p_m\approx 250\ \text{MeV}/c$
and $\gamma_1\approx 30^{\circ}$.  Figure\ \ref{fig:p13} shows the
cross section as a function of $p_{13}$ for the measurements at
$q$=445\ MeV/$c$, where the detection volume extends to
$p_{13}=0\ \text{MeV}/c$.  The acceptance in $p_m$ has been limited to
$380<p_m<400\ \text{MeV}/c$ to ensure complete coverage of the
detection volume for the entire domain $0<p_{13}<320\ \text{MeV}/c$. 
This also limits $p^\prime_1$ to the range 380--430\ MeV/$c$.  The
range in $p_{13}$ is therefore mainly due to angular variations
between the undetected neutron and the forward proton.  A typical
final-state configuration is shown in the inset of Fig.\ \ref{fig:p13}.

The cross section data presented in Fig.\ \ref{fig:p13} show an
increase as $p_{13}$ approaches 0\ MeV/$c$.  This trend is well
reproduced by the continuum Faddeev calculations which include
rescattering among the three outgoing nucleons.

\begin{figure}
    \epsfxsize=3.375in
    \centerline{\epsffile{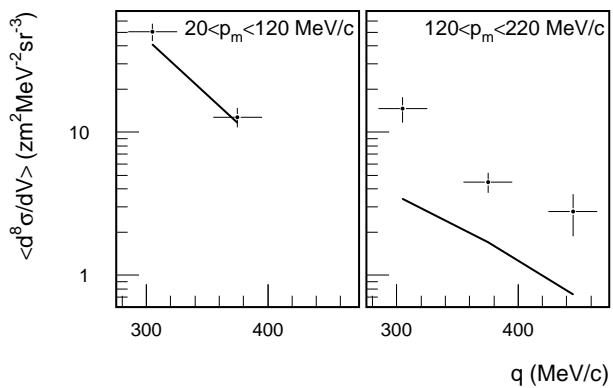}}
    \caption{
		Averaged cross section for three values of the virtual-photon
		momentum at an energy transfer of 220\ MeV.  The cross
		sections were averaged over $10^{\circ}<\gamma_1 <25^{\circ}$
		and the indicated missing-momentum intervals.  The horizontal
		error bars indicate the range covered in $q$ by the acceptance
		of the spectrometer.  The curves represent the results of
		continuum Faddeev calculations with a one-body hadronic
		current.
		\label{fig:qdep}
	}
\end{figure}

Figure\ \ref{fig:qdep} shows the dependence of the cross section on the
three-momentum transfer $q$.  As the cross section is primarily
determined by the momentum of the neutron, the data are presented for
two regions in $p_m$.  The region above $p_m$=220\ MeV/$c$ is
disregarded in this discussion as `FSI configurations' occur in this
domain and the cross section is affected differently at various 
values of $q$.

For $20<p_m<120\ \text{MeV}/c$ the reaction can be considered as a
direct process, leaving the spectator neutron `at rest' in the final
state.  The measured cross section decreases by a factor of four
between $q$=305 and 375\ MeV/$c$.  The agreement in size and slope
between the data and the continuum Faddeev calculations performed with
a one-body hadronic current, suggests that in this domain the cross
section is dominated by knockout of correlated proton pairs.  For
$120<p_m<220\ \text{MeV}/c$ theory underestimates the data for all
three $q$ values by a factor of five.

In conclusion, differential cross sections for the
$^{3}\text{He}(e,e^{\prime}pp)$ reaction were measured at
$\omega$=220\ MeV and $q$=305, 375, and 445\ MeV/$c$ with good
statistical accuracy.  The measured cross section decreases roughly
exponentially as a function of $p_m$ in the domain from 10 to
350\ MeV/$c$.  Continuum Faddeev calculations using the Bonn-B $NN$
potential, which include rescattering of the outgoing nucleons to all
orders and also take meson-exchange currents into account, describe the
data reasonably well for $p_m<100\ \text{MeV}/c$.  However, an
increasing discrepancy up to a factor three is observed at larger $p_m$
values.  In this domain meson-exchange currents account for 40\% of
the calculated cross section and also intermediate $\Delta$
excitation, especially in the $pn$ system, is expected to contribute
strongly to the cross section in this domain.  Comprehensive treatment
of intermediate $\Delta$ excitation in the calculations is needed for
a detailed interpretation of the data in this high $p_m$ region. 
Rescattering of the nucleons in the final state influences the cross
section in specific kinematic orientations; within such a subset of
the data, the observed trend is well reproduced by the results of the
continuum Faddeev calculations.

At small $p_m$ values the dependence of the cross section on $q$ is
indicative for direct knockout of two protons by a virtual photon. 
The fair agreement between the data and calculations based on a
one-body hadronic current indicates that in this domain the cross
section mainly originates from knockout of two correlated protons. 
This opens the opportunity to exploit this low $p_m$ domain for a
detailed study of $pp$ correlations in few-nucleon systems.

This work is part of the research program of the Foundation for
Fundamental Research on Matter (FOM) and was sponsored by the
Stichting Nationale Computerfaciliteiten (National Computing
Facilities Foundation, NCF) for the use of supercomputer facilities. 
Both organisations are financially supported by the Netherlands
Organisation for Scientific Research (NWO).  The support of the
Science and Technology Cooperation Germany-Poland, the Polish
Committee for Scientific Research (grant No.  2P03B03914), and the US
Department of Energy is gratefully acknowledged.  Part of the
calculations have been performed on the Cray T90 and T3E of the John
von Neumann Institute for Computing, J\"ulich, Germany.

\end{document}